# Weak Localization in Few-Layer Black Phosphorus

Yuchen Du, Adam T. Neal, Hong Zhou and Peide D. Ye[*]

School of Electrical and Computer Engineering and Birck Nanotechnology Center,

Purdue University, West Lafayette, Indiana 47907, United States

*E-mail: yep@purdue.edu

Abstract

We have conducted a comprehensive investigation into the magneto-transport properties of few-layer black phosphorus in terms of phase coherence length, phase coherence time, and mobility *via* weak localization measurement and Hall-effect measurement. We present magnetoresistance data showing the weak localization effect in bare p-type few-layer black phosphorus and reveal its strong dependence on temperature and carrier concentration. The measured weak localization agrees well with the Hikami-Larkin-Nagaoka model and the extracted phase coherence length of 104 nm at 350 mK, decreasing as $\sim T^{-0.51 \pm 0.05}$ with increased temperature. Weak localization measurement allows us to qualitatively probe the temperature-dependent phase coherence time $\tau_\phi$, which is in agreement with the theory of carrier interaction in the diffusive regime. We also observe the universal conductance fluctuation phenomenon in few-layer black phosphorus within moderate magnetic field and low temperature regime.

Keywords: black phosphorus, weak localization, phase coherence length, Hall mobility

Introduction

Black phosphorus (BP) is a stable phosphorus allotrope at room temperature [1, 2]. The structure of BP is that of a layered material in which individual atomic layers are stacked together by van der Waals interactions [3-5]. The discovery of BP can be dated back a century ago, where the physical properties of the bulk crystal were extensively studied due to its favorable electronic characteristics, including a direct band gap of 0.3 eV [1, 3, 5], a high hole mobility of up to $10^4$ cm$^2$/Vs at low temperature [6], and anisotropic transport in the 2D plane [3, 7, 8]. Similar to graphene, bulk BP can be isolated by the standard mechanical exfoliation method, where the single layer BP, known as "phosphorene" has been recently demonstrated [9]. The importance of developing single or few-layer BP has led the society to the emergence of a new paradigm of condensed-matter physics, where the quantum phenomena such as Quantum Hall effect and Shubnikov-de Haas oscillations have been now observed in high-quality bare or passivated black phosphorus films [10-15]. Much more research has be reported on this narrow-bandgap 2D semiconductor in terms of its electronic, optical, thermal and mechanical properties [16-23]. In this paper, we report on the magneto-transport studies of few-layer BP films, focusing on weak localization and its temperature and carrier concentration dependence. The weak localization effect can be fitted by the Hikami-Larkin-Nagaoka (HLN) model where the temperature dependence of the phase coherence length has demonstrated a power-law behavior of ~ $T^{-0.51 \pm 0.05}$. The maximum observed phase coherence length is 104 nm at 350 mK on bare BP films, elucidating the possibility to explore phase coherent electronic devices with the state-of-the-art nanofabrication techniques. Additionally, universal conductance fluctuation (UCF) effect has also been observed in our few-layer BP device. The study of Hall mobility with different temperatures and carrier concentration has also been performed in our experiment. A Hall mobility of 407 cm$^2$/Vs from bare few-layer BP film, at a carrier concentration of $9.66 \times 10^{12}$ cm$^{-2}$, has been observed at 350 mK. Due to the instability of BP films in ambient, the thinnest film so far we fabricated successfully and systematically characterized at low temperatures in this work is 4.54 nm.

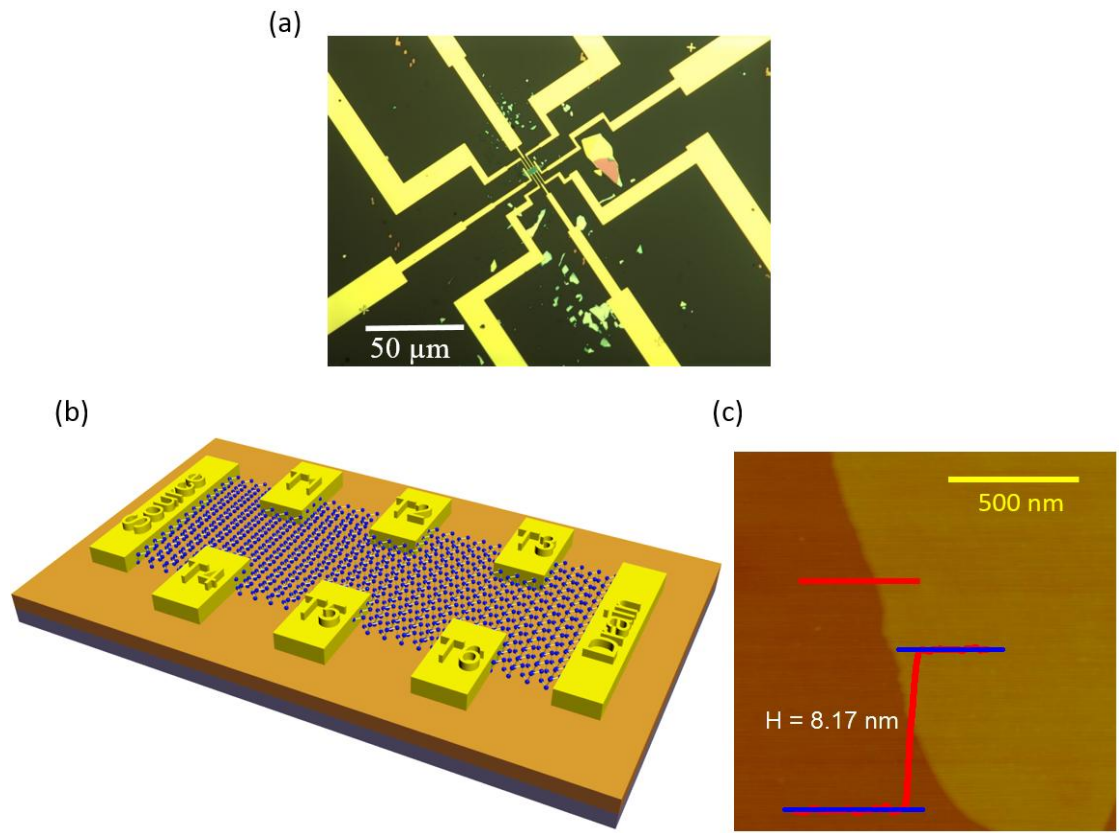

Figure 1: (a) Optical image of fabricated device. (b) Schematic view of device configurations. A p$^{++}$ silicon wafer capped with 90 nm SiO$_2$ was used as the global gate and gate dielectric, respectively. Few-layer BP film was exfoliated from bulk BP crystals. Ni/Au was evaporated as the contacts. A magnitude of 1 μA testing current has been given from the source to the drain. The longitude resistance $R_{xx}$ has been recorded between T1 and T3, and Hall resistance $R_{xy}$ has been measured between T1 and T4. Source-drain distance is designed as 7.0 μm, longitude distance T1-T3 is 3.5 μm, and Hall distance T1-T4 is 2.5 μm. (c) Atomic force microscopy image of this few-layer BP film with a measured thickness of 8.17 nm.

Discussion

The optical and schematic view of the few-layer BP Hall bar device is illustrated in figure 1(a) and (b), and a magnitude of 1 µA input current has been sent from the source to the drain, which is large enough to minimize measurement noise, but not too much to induce current heating at low temperature with several kilo-ohms resistance. The longitude resistance $R_{xx}$ is recorded between T1 and T3, and Hall resistance $R_{xy}$ is measured between T1 and T4 *via* magneto-transport measurements. The thickness of film for this device is 8.17 nm as shown in figure 1(c), measured by the atomic force microscopy system (AFM). Weak localization and Hall measurements were carried out in a He$^3$ cryostat with a superconducting magnet using a Stanford Research 830 lock-in amplifier.

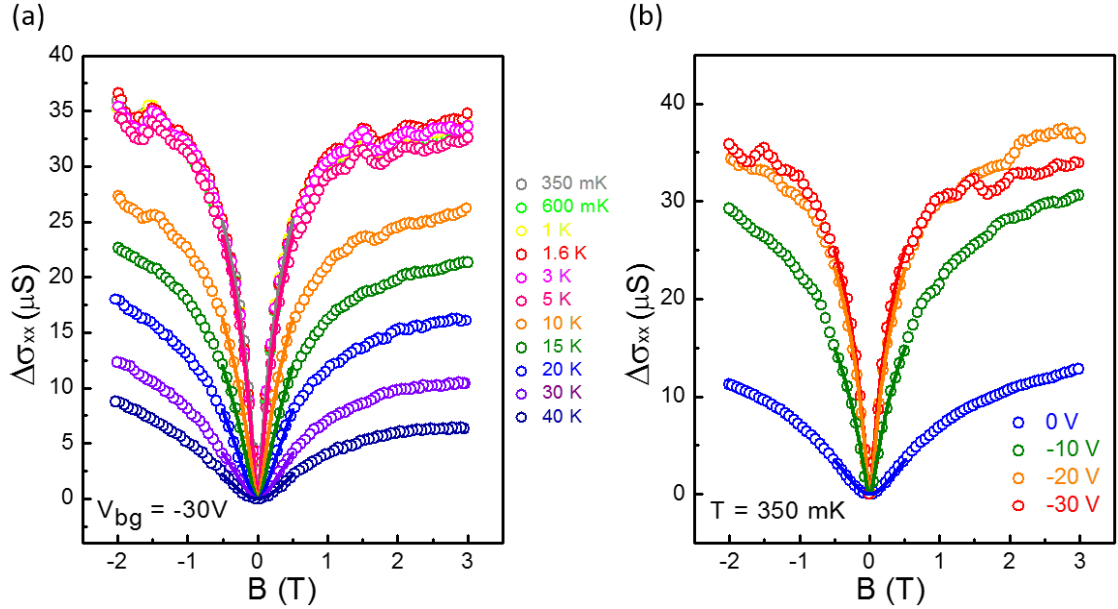

Figure 2: Magneto-conductivity measurements of weak localization (a) At constant back gate bias of -30 V for various temperatures from the base temperature of 350 mK up to 40 K. (b) At base temperature of 350 mK for back gate voltages of 0 V, -10 V, -20 V, and -30 V. The solid lines are fitting curves from HLN model within -500 mT and 500 mT.

Weak localization in few-layer BP

The weak localization effect originates from the constructive interference of backscattered electronic wave functions which increases the probability of scattering an electron. This phenomenon manifests itself by a positive correction to resistivity (negative correction to the conductivity) at small magnetic fields. The applied vertical magnetic field induces an additional phase difference to break the constructive interference, leading to negative magneto-resistance and positive magneto-conductance [24, 25]. The magneto-conductivity is calculated from the measured longitude resistivity $\rho_{xx}$ and Hall resistivity $\rho_{xy}$ by matrix inversion, and we have observed a positive magneto-conductivity at zero magnetic field for a variety of temperatures and carrier concentrations, which is a standard characteristic of weak localization effect. Magneto-conductivity measurements of weak localization with different temperatures and carrier concentrations have been shown in figure 2. As expected, the weak localization is strongly temperature dependent and demonstrates a deep dip on the base temperature of 350 mK, which is plotted in figure 2(a). In addition, within the low temperature regime, from 350 mK to 5 K, we did not observe a strong temperature dependence, where the weak localization dips almost coincide with one another. However, the effect dies out when the temperature increases to 40 K, in which the weak localization is suppressed when the phase coherence length shrinks at higher temperatures. Magneto-conductivity measurements of weak localization at a temperature of 350 mK for various of back gate voltages have also been illustrated in figure 2(b). We find that the weak localization effect is strongly gate-tunable, where the high hole concentration in few-layer BP films, thus a high hole mobility, yields a larger correction to the magneto-conductivity at $V_{bg}$

= -30 V. The cause of gate dependent weak localization behavior can also be attributed to the variation of phase coherence length with various carrier concentrations. A detailed examination on phase coherence length and phase coherence time, which would be discussed in the later parts of this paper, helps us understand the temperature and carrier concentration dependent characteristics of weak localization in few-layer BP films.

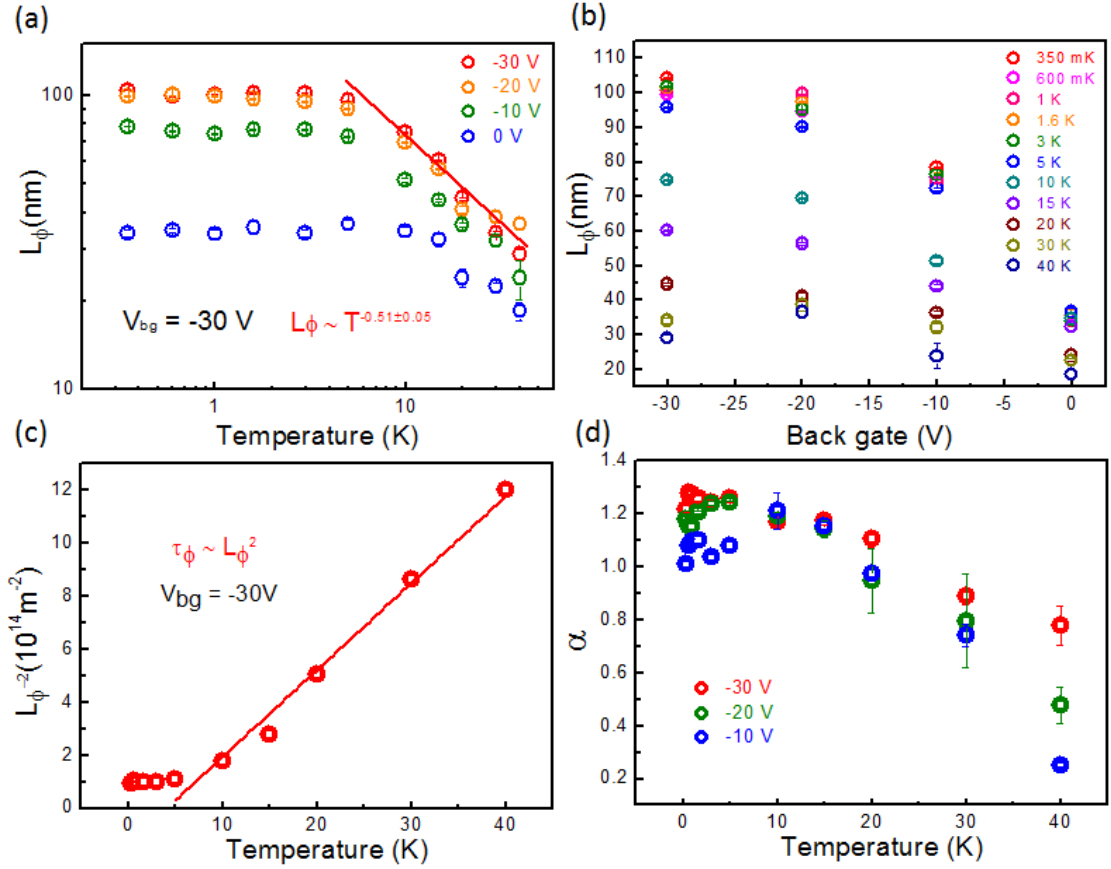

Figure 3: (a) Phase coherence length varies with temperature at different back gate voltages in log-log scale. The temperature dependence of the phase coherence length has demonstrated a power-law behavior of $\sim T^{-0.51\pm0.05}$ at the condition of $V_{bg}$ = -30 V. The red solid line is power-law cure fitting line. (b) Phase coherence length varies with back gate voltage at different temperatures. (c) Phase coherence length, $L_\phi^{-2}$ varies with different temperatures at a constant back gate voltage of -30 V. The phase coherence time is proportional to the $L_\phi^2$, with a relation of $\tau_\phi \sim L_\phi^2$. The red solid line is a linear fitting curve from 5 K to 40 K. (d) Empirical fitting parameter, α, varies with temperature at different back gate voltages. Error bars are determined from HLN model fitting.

Parameter extraction from HLN model

To gain insight into the phase coherence length, phase coherence time, and empirical fitting parameter, we fit the low-field portion of the magneto-conductance curves using the HLN model [26], where the equation has been widely applied to analyze $\Delta\sigma(B)$ due to the weak localization effect:

$$\Delta\sigma(B) = \sigma(B) - \sigma(B=0) = \alpha \frac{e^2}{2\pi^2 \hbar} F(\frac{B}{B_\phi})$$

$$F(Z) = \psi(\frac{1}{2} + \frac{1}{Z}) + ln(Z), B_\phi = \frac{\hbar}{4eL_\phi^2}$$

where $\psi$ is the digamma function, $L_\phi$ is the phase coherence length, $e$ is the electronic charge, $\hbar$ is the reduced Planck's constant, $B$ is the magnetic field, and $\alpha$ is the empirical fitting parameter. We should note that, the implied spin degeneracy of $n_s = 2$, and valley degeneracy of $n_v = 2$ in few-layer BP should also be considered in empirical fitting parameter. In addition, we have limited our curve fitting to magnetic fields in the range of -500 mT to 500 mT to avoid effects of the classical magnetoresistance at low carrier density. Phase coherence length vs. temperature at various back gate biases has been shown in figure 3(a). Similar to the magneto-conductivity measurement, in the low temperature regime from 350 mK to 5 K, the value of phase coherence length does not change significantly with temperatures, which may be attributed to the fact that the phase coherence length is maintained over the length which is induced by the defects in material [27-29]. The maximum phase coherence length extracted from the low-$B$ field portion is ~104 nm at 350 mK and -30 V back gate bias. As we increase the temperature from 5 K to 40 K, the temperature dependence of the phase coherence length demonstrates a strong power-law behavior of ~ $T^{-\gamma}$. In order to accurately extract the power exponent $\gamma$, we have fitted the temperature dependent curve from 5 K to 40 K, shown in figure 3(a) as a solid red line. At -30 V back gate bias, the phase coherence length decays with temperature as ~$T^{-0.51\pm0.05}$, which matches the observation from previous studies that electron-electron or hole-hole scattering would give $L_\phi$ proportional to $T^{-0.5}$ in a 2D system [27, 30]. It is interesting to note that, at different back gate biases, such power-law fittings give $L_\phi$ proportional to $T^{-0.47\pm0.04}$, $T^{-0.49\pm0.02}$, and $T^{-0.46\pm0.07}$ for $V_{bg}$ = -20 V, -10 V, and 0 V, respectively. The consistent values of power exponent $\gamma$ not only explain well the fundamental mechanism of hole-hole scattering in 2D BP system, but also show the reliability of all measurements as well. The relationship between phase coherence length and different back gate bias has been shown in figure 3(b). The conductivity dip at zero magnetic field gets deeper as the gate voltage has been increased from 0 V to -30 V, resulting from an enhancement of phase coherence length $L_\phi$ with increasing carrier concentration thus hole mobility in the few-layer BP film. Phase coherence time, which is another important characteristic in weak localization effect, is also unveiled in our study. The phase coherence length is related to the phase coherence time by the diffusion coefficient as $L_\phi=\sqrt{D\tau_\phi}$, where $\tau_\phi$ is phase coherence time, and $D$ is diffusion coefficient. Meanwhile, the diffusion coefficient is found from the relation $D = \frac{\sigma}{e^2 g_{2D}(E_F)} = \frac{\sigma\pi\hbar^2}{m^*e^2}$ where $\sigma$ is the conductivity in channel, and $m^*$ the effective mass of carriers in the few-layer BP. Notably, we should notice that, due to the anisotropic transport property in BP, the effective mass of carriers cannot be easily evaluated from device perspectives, however, instead of calculating the exact value of phase coherence time, quantitative behavior analysis of phase coherence time can be realized in our research. $L_\phi^{-2}$, which is directly proportional to $\tau_\phi^{-1}$, versus temperature has been plotted in figure 3(c). The linear dependent increase of the phase coherence time with temperature agrees well with the prediction of the theory of hole-hole interaction in the diffusive regime, showing conventional behavior of carrier interference [32]. Empirical fitting parameter $\alpha$ varies with temperature and carrier concentration and has also been plotted in figure 3(d), where $\alpha$ is found to be close to theoretical value of 1, and the valley degeneracy $n_v = 2$ is also applied in this empirical fitting parameter.

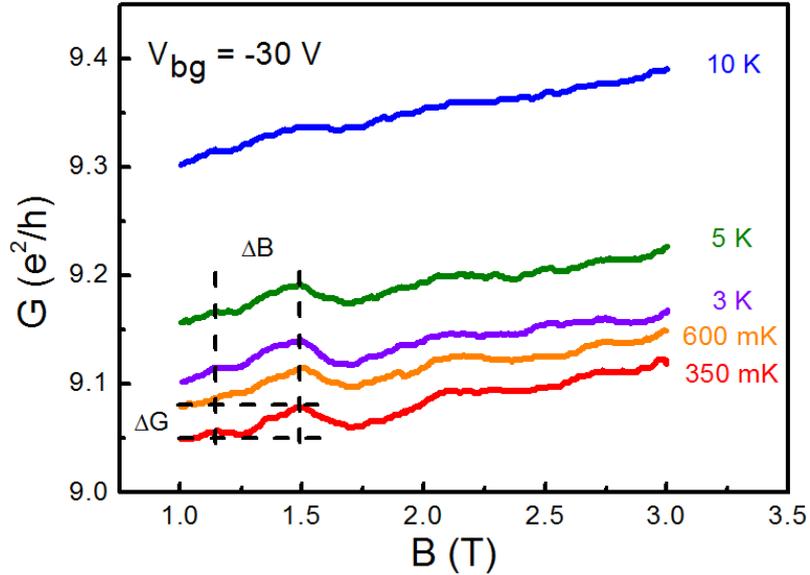

Figure 4. Magneto-conductance fluctuation in few-layer BP at fixed temperatures of T = 350 mK, 600 mK, 3 K, 5 K, and 10 K. The measurements are taken at a back gate modulation of -30 V.

## UCF in few-layer BP

UCF is a quantum interference effect of diffusive charge carriers, observed commonly in the mesoscopic systems of semiconductors, metals, and graphene [33, 34]. The mesoscopic nature or finite size of a weakly disordered sample results in the loss of the self-averaging of its physical properties. Applying magnetic field in semiconductors varies the phase of the wave function of a charge carrier, where the magnitude and interval of conductance fluctuation are closely related to the phase coherence length [33,34]. Interestingly, the UCF effect is observed in our few-layer BP sample within the moderate magnetic field and low temperature regime, and this phenomenon has been confirmed by repeatable measurements from switching different terminal legs in Hall bar structure. In figure 4, we demonstrate the magneto-conductance traces of the order of $e^2/h$ at fixed back gate modulation of -30 V for the few-layer BP film at different temperatures. Device results confirm that the existing of semiconducting coherent phase around certain defects in few-layer BP. The conductance is gradually decreasing as temperature drops from 10 K down to 350 mK, where a negative temperature derivative of $R_{xx}$ ($dR_{xx}/dT$) is obtained. In addition, it is clear to see that the UCF is strongly temperature dependent, where the fluctuation remains consistent within the low temperatures from 350 mK up to 5 K. However, it decays quickly as the temperature passes the 10 K point, being consistent with the temperature dependent coherence length measurement depicted in figure 3. The characteristic interval for magneto-conductance fluctuation $\Delta B$, with a value of 0.38 T for this few-layer BP film at a base temperature of 350 mK, is robust, and persisting to the temperatures as high as 5 K. From $\Delta B=(h/e)/(\pi r)^2$, we can also estimate the mesoscopic "defect" or potential fluctuation has the dimension of ~120 nm, again being consistent with the coherent length in figure 3. The amplitude of magneto-conductance fluctuation, $\Delta G$ is 0.027 $e^2/h$ at T = 350 mK for the few-layer sample, which keeps almost the same within temperature regime of 350 mK to 5 K. Furthermore, both magnitude and interval of conductance fluctuation die out rapidly as the temperature increases, since the phase coherence length decays with increase of temperature with the power law.

Magneto-conductance fluctuation in BP confirms the picture that the magnitude and interval of conductance fluctuation is closely associated to the magnitude of phase coherence length, where the UCF is suppressed as the phase coherence length is smaller than 70 nm.

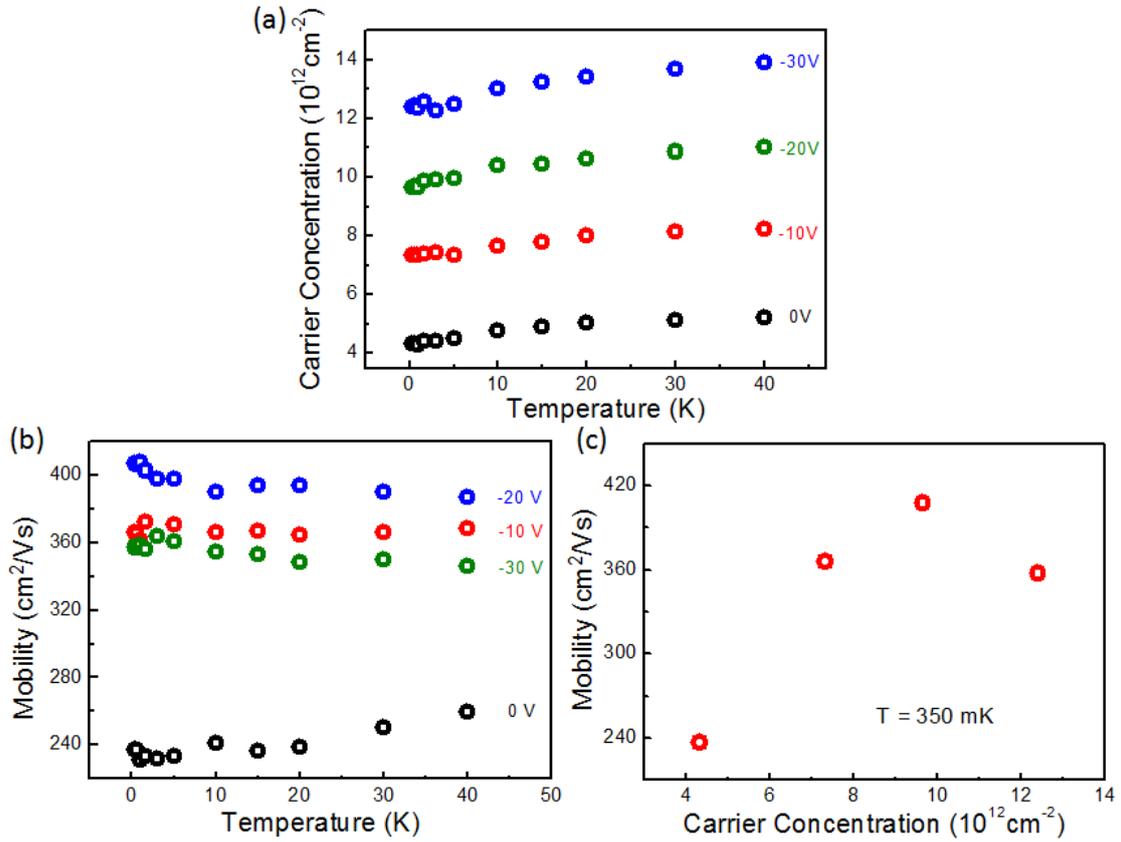

Figure 5: (a) Carrier concentration varies with temperatures at different back gate biases. Four different back gate voltages of 0 V, -10 V, -20 V, and -30 V are applied in the measurements. (b) Hall mobility as a function of temperature at different back gate biases. (c) Hall mobility versus carrier concentrations at the base temperature of 350 mK. Four different 2D carrier densities are corresponding to four different back gate voltages of 0 V, -10 V, -20 V, and -30 V, respectively.

Temperature and carrier dependent Hall mobility

Temperature dependence of carrier concentration and mobility are examined in this work to uncover the limitation of mobility in bare few-layer BP without passivation. We have performed Hall-effect measurements on single-gated BP supported on $SiO_2$ from 40 K down to the base temperature of 350 mK. The two dimensional carrier density $n_{2D}$ is determined from $n_{2D} = \frac{B}{e\rho_{xy}}$. Hall resistance $R_{xy}$ or Hall resistivity $\rho_{xy}$ follows a linear dependence on the magnetic field $B$. As depicted in figure 5(a), the intrinsic carrier concentration of $4.33 \times 10^{12}$ cm$^{-2}$ in few-layer BP is recorded at the zero back gate bias. As we electrostatically p-dope the few-layer BP *via* an application of -30 V back gate bias, the hole carrier density has been pushed up to its maximum value of $n_{2D} = 1.24 \times 10^{13}$ cm$^{-2}$ at the base temperature. Hall mobility, obtained from $\mu_H = \frac{L}{W} \frac{1}{R_{xx} n_{2D} e}$, where $L$ is channel length, $W$ is channel width, and $R_{xx}$ is longitude resistance, has been systematically studied in this experiment. The Hall mobilities for different back gate bias in few-layer BP as a function of temperature are shown in figure 5(b). The maximum Hall mobility occurs at the -20 V back gate bias, with a magnitude of 407 cm$^2$/Vs. For four different bias conditions, they all show a similar trend that the mobility is saturated at a temperature range of 40 K down to 350 mK, which matches the previous report of mobility behaviors of an 8 nm thick sample at low temperatures [10]. We also note that in all temperature ranges, the Hall mobility increases as the gate-induced carrier density becomes larger, however, as the back gate bias exceeds -30 V, the Hall mobility of few-layer BP starts to be reduced. Hall mobility versus carrier concentration at the base temperature of 350 mK are plotted in figure 5(c). The mobility has been enhanced from 237 cm$^2$/Vs to 407 cm$^2$/Vs, corresponding to an increase of carrier concentration from $4.33 \times 10^{12}$ cm$^{-2}$ to $9.66 \times 10^{12}$ cm$^{-2}$. On the other hand, it drops to 357 cm$^2$/Vs as carrier concentration touches $1.24 \times 10^{13}$ cm$^{-2}$. The observed concentration-dependent phenomenon indicates that different scattering mechanisms are limiting the Hall mobility at different carrier concentration regions. For intermediate carrier concentration, $n_{2D} < 9.66 \times 10^{12}$ cm$^{-2}$, the observed trend suggests that scattering from charged impurities limits the mobility [35]. As carrier concentration increases from $4.33 \times 10^{12}$ cm$^{-2}$ to $9.66 \times 10^{12}$ cm$^{-2}$, the reduced scattering in the BP points to the diminished disorder potential as a result of screening by free charge carriers [10,14, 36]. On the other hand, in higher carrier concentration regime when the hole carrier density exceeds $9.66 \times 10^{12}$ cm$^{-2}$, the Hall mobility starts to drop from the maximum value of 407 cm$^2$/Vs down to 357 cm$^2$/Vs. It can be ascribed to the enhanced random charged impurity scattering at BP/SiO$_2$ interface when the very high gate bias is applied [10-14, 36].

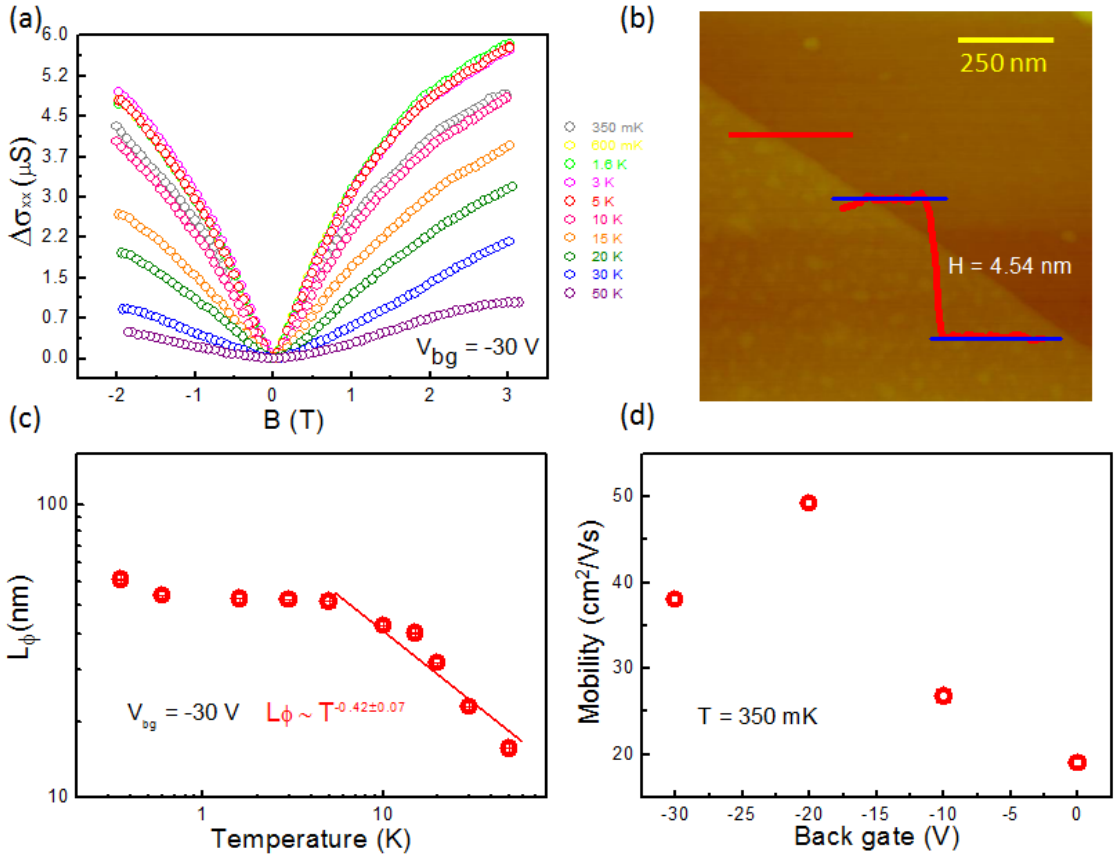

Figure 6: (a) Magneto-conductivity measurements of few-layer phosphorene weak localization at a constant back gate bias of -30 V for various temperatures. (b) Atomic force microscopy image of this few-layer phosphorene film with a measured thickness of 4.54 nm. (c) Phase coherence length of few-layer phosphorene varies with temperatures at a constant back gate voltage of -30 V. The temperature dependence of the phase coherence length has demonstrated a power-law behavior of ~$T^{-0.42 \pm 0.07}$ at the condition of $V_{bg}$ = -30 V. The red solid line is power-law cure fitting line, and error bars are determined from HLN model fitting. (d)

Hall mobility of few-layer phosphorene versus back gate modulations at the base temperature of 350 mK.

## Weak localization in thinner layer BP

Since BP or phosphorene are not very stable in air, it is difficult to fabricate and study the magneto-transport properties of single-layer or bi-layer phosphorene films without the development of the sophisticated passivation techniques [37,38]. The thinnest bare few-layer BP samples we have successfully fabricated and studied in this work is 4.54 nm thick. Magneto-conductivity measurements of weak localization in this ultrathin BP film at a constant back gate bias of -30 V for various temperatures from 350 mK up to 50 K are plotted in figure 6(a). Similar to the 8 nm BP flake, a strong temperature dependent behavior of the weak localization effect are observed, where the shape of weak localization fades out when the sample temperature heats up from 5 K to 50 K. However, the magnitude of the weak localization dip is smaller in 4nm sample as compared to 8nm BP one. The differential magneto-conductivity is also described for 2D weak localization by the HLN equation to analysis the phase coherence length within the fitting limitation of 500 mT to -500 mT. The extracted phase coherence length at -30 V back gate voltage for different temperatures is unfolded in figure 6(b). The measured phase coherence length ranges from 49.1 nm at 600 mK to 14.7 nm at 50 K for $V_{bg}$ = -30 V, and the phase coherence length varies with temperature as a power-law behavior of ~ $T^{-0.42 \pm 0.07}$. Note that the values of phase coherence length in 4nm BP film are roughly half of those in 8nm BP films, as depicted in figure 3(a). We believe this reduction of phase coherence length is a result of decreased mobility for thinner BP film. The carrier concentration in channel is mainly controlled by the back gate voltage, where the hole densities in thick and thin BP samples are $1.24 \times 10^{13}$ cm$^{-2}$ and $1.01 \times 10^{13}$ cm$^{-2}$, respectively, at the similar range with -30 V back gate bias. As presented in figure 6(c), Hall mobility of ultrathin BP film is 49.2 cm$^2$/Vs at the back gate of -20 V, significantly smaller than that from the thick BP film. This lower Hall mobility confirms the degradation of film quality during fabrication process in ambient atmosphere. The defects and traps at BP/SiO$_2$ interface could also have more negative effects on thinner BP film. It is also worth mentioning that there is no obvious UCF effect observed in 4.54 nm ultrathin film, since the phase coherence length is significantly reduced. Because the width of our flake is micrometers scale, the fringe current flow in the real device would partly average out the anisotropy [9,17], and lattice orientation induced anisotropic mobility is not included in our above discussions.

Conclusion

In summary, we have characterized the phase coherence length and Hall mobility in few-layer BP films *via* Hall and weak localization measurements. The detailed analysis of weak localization indicates a maximum phase coherence length of 104 nm at base temperature of 350 mK, and the effect also demonstrates a strong temperature dependent power-law behavior of ~ $T^{-0.51 \pm 0.05}$. In 2D semiconducting materials, the UCF effect has been observed in 8.17 nm thick BP film, at the moderate magnetic field and low temperature regime. The observed density dependent mobility trend indicates that different scattering mechanisms are limiting the Hall mobility in BP films. Much lower phase coherence length and mobility values are obtained in ultrathin BP films due to its instability in ambient atmosphere.

Acknowledgements

This material is based upon work partly supported by NSF under Grant ECCS-1449270, AFOSR/NSF EFRI 2-DARE Grant No. 1433459-EFMA, and ARO under Grant W911NF-14-1-0572. The low temperature measurements were performed at the National High Magnetic Field Laboratory (NHMFL), which is supported by National Science Foundation Cooperative Agreement No. DMR-1157490, the State of Florida, and the U.S. Department of Energy. The authors would like to thank Y. Xu, Dr. J. Tian for the valuable discussions, and T. Murphy, J.-H. Park, and G. Jones for experimental assistance.